\documentclass[twocolumn,showpacs,preprintnumbers,floatfix,amsmath,amssymb,prd]{revtex4}

\usepackage{dcolumn}
\usepackage{bm}
\usepackage{color}
\usepackage{graphicx}

\def\lsim{\lesssim}

\def\beq{\begin{equation}}
\def\eeq{\end{equation}}
\def\be{\begin{eqnarray}}
\def\ee{\end{eqnarray}}

\begin{document}
\title{Electroweak nuclear response in quasi-elastic regime}

\author{Omar Benhar$^{1,2}$}
\author{Pietro Coletti$^{2}$}
\author{Davide Meloni$^{3}$}
\affiliation
{$^1$INFN, Sezione di Roma, I-00185 Roma, Italy \\
$^2$Dipartimento di Fisica, ``Sapienza'' Universit\`a di Roma, I-00185 Roma, Italy \\
$^3$Institut f\"ur Theoretische Physik und Astrophysik \\
Universit\"at W\"urzburg, D-97974   W\"urzburg, Germany
 }

\date{\today}

\begin{abstract}
The availability of the double-differential charged-current neutrino cross section, measured by the 
MiniBooNE collaboration using a carbon target, allows for a systematic 
comparison of nuclear effects in quasi-elastic electron and neutrino scattering. The results of our study, based 
on the impulse approximation scheme and a state-of-the-art model of the nuclear spectral functions, 
suggest that the electron cross section and the flux averaged neutrino cross sections, corresponding to the 
same target and comparable kinematical conditions, cannot  be described within the same theoretical approach
using the value of the nucleon axial mass obtained from deuterium measurements. We analyze the 
assumptions underlying the treatment of electron scattering data, and argue that the description
of neutrino data will require a new {\em paradigm}, suitable for application to processes in which the lepton kinematics
is not fully determined.

\end{abstract}



\maketitle


The data set of Charged Current Quasi Elastic (CCQE) events recently released by the MiniBooNE 
collaboration \cite{BooNECCQE}  provides an unprecedented opportunity to carry out a systematic
study of the double differential cross section of the process, 
\beq
\nu_\mu + ^{12}\mkern -5mu C \rightarrow \mu^- + X \ ,
\eeq
integrated over the neutrino flux.  Comparison between the results of theoretical calculations and data
may provide valuable new information on nuclear effects, whose quantitative understanding 
is critical to the analysis of neutrino oscillation experiments, as well as on the elementary interaction 
vertex. 

The charged current elastic neutrino-nucleon process is described in terms of three 
form factors. The vector form factors $F_1(Q^2)$ and $F_2(Q^2)$ ($Q^2= -q^2$, $q$ being the four-momentum 
transfer) have been precisely measured, 
up to large values of $Q^2$, in electron-proton and electron-deuteron scattering experiments 
(for a recent review, see, e.g., Ref.\cite{VFF}). The $Q^2$-dependence of the axial form 
factor $F_A(Q^2)$, whose value at $Q^2=0$ can be extracted from neutron $\beta$-decay
measurements, is  generally assumed to be of dipole form and parametrized in terms of  the so 
called axial mass $M_A$:
\beq
F_A(Q^2) = g_A \ \left( 1 + Q^2/M_A^2 \right)^{-2} \ .
\eeq
The world average of the measured values of the axial mass, mostly obtained 
using deuterium targets, turns out to be  $M_A = 1.03 \pm 0.02$ GeV \cite{bernard,bodek,nomad}, 
while analyses carried out by the K2K \cite{K2K} and MiniBooNE \cite{BOONE} 
collaborations using oxygen and carbon targets, respectively, yield   
$M_A \sim 1.2 \div 1.35 \ {\rm GeV}$.
 
It would be tempting  to interpret the large value of $M_A$ reported by MiniBoonNE
and K2K as an {\em effective} axial mass, modified by nuclear effects not included in the
Fermi gas model employed in data analysis. However, most existing models of nuclear 
effects (for recent reviews see Ref.\cite{nuint09}) fail to support this explanation,  
suggested  by the authors of Ref. \cite{BOONE}, a prominent exception being the model of Ref. \cite{martini}. 

Obviously, a fully quantitative description of the electron-scattering cross section, driven by the known 
vector form factors, is a prerequisite for the understanding of the axial vector contribution to the 
CCQE neutrino-nucleus cross section. 
 
Over the past two decades, the availability of a large body of experimental data has triggered 
the development of advanced theoretical descriptions of the nuclear electromagnetic response. 
The underlying scheme, based on nuclear many-body theory and realistic nuclear hamiltonians, 
relies on the premises that i) the lepton kinematics is fully determined and ii) the elementary 
interaction vertex can be extracted from measured proton and deuteron cross sections. 

The above {\em paradigm} has been successfully applied to explain the electron-nucleus cross section 
in a variety of kinematical regimes (for a recent review of the quasi-elastic sector see Ref. \cite{RMP}). However,  in view of the 
uncertainties associated with the energy of the incoming beam, the identification of the reaction 
mechanisms and the determination of the interaction vertex, its extension to the case of neutrino 
scattering may not be straightforward.

In this work we compare 
theoretical results obtained from the approach described in Refs. \cite{PRD,NPA} to the measured CCQE cross 
sections of Ref.\cite{BooNECCQE}, discuss the differences involved in the  analyses of electron and neutrino-nucleus 
scattering, and argue that modeling neutrino interactions may require a paradigm shift.

\begin{figure}[h!]
\includegraphics[width=0.75\columnwidth]{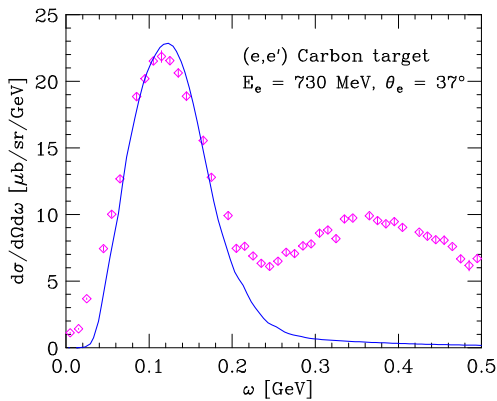}
\caption{(color online). Inclusive electron-carbon cross section at beam energy $E_e=$ 730 MeV and electron scattering 
angle $\theta_e=37^\circ$, plotted as a function of the energy loss $\omega$. The data points are taken from 
Ref. \cite{12C_ee}.}
\label{xsec_ee}
\end{figure}

Electron-nucleus scattering cross sections are usually analyzed at fixed beam energy, $E_e$, and 
electron scattering angle $\theta_e$, as a function of the electron energy loss $\omega$.
As an example, Fig. \ref{xsec_ee} shows the double differential cross section of the process
\beq
e + ^{12}\mkern -5mu C \rightarrow e^\prime + X \ ,
\eeq
at $E_e = 730$ MeV and $\theta_e = 37^\circ$, measured at MIT-Bates~\cite{12C_ee}.
The peak corresponding to quasi-elastic (QE) scattering and the bump at larger $\omega$, associated 
with excitation of the $\Delta$-resonance, are clearly visible and well separated. 
Note that the three-momentum transfer $|{\bf q}|$ turns out to be nearly constant, its variation over the range 
shown in the figure being $\lsim$ 5\%. As a consequence, the cross section of Fig.\ref{xsec_ee} can be 
readily related to the linear  response of the target nucleus to a probe delivering momentum ${\bf q}$ and 
energy $\omega$, defined as
\beq
\label{Sqw}
 S({\bf q},\omega) = \sum_n | \langle n | \sum_{{\bf k}} a^\dagger_{{\bf k}+{\bf q}} a_{{\bf k}} | 0 \rangle |^2
 \delta(\omega  + E_0 - E_n)\ .
 \eeq
In the above equation, $| 0 \rangle$ and $| n \rangle$ represent the target initial and final states, having 
energy $E_0$ and $E_n$, respectively, while  $a^\dagger_{{\bf k} + {\bf q}}$ and $a_{{\bf k}}$ are the nucleon 
creation and annihilation operators.

 In addition, the magnitude of the momentum transfer,  $|{\bf q}|~\sim$~450~MeV, 
is large enough to make the impulse approximation (IA) scheme, in which Eq.(\ref{Sqw}) reduces 
to \cite{NUINT07}
\beq
S_{IA}({\bf q},\omega) = \int d^3k \ dE P_h({\bf k},E) P_p({\bf k}+{\bf q},\omega-E) \ ,
\label{S:IA}
\eeq
safely applicable \cite{shape}. In  Eq.(\ref{S:IA}), $P_h(~{\bf k}~,~E~)$ and $P_p({\bf k}~+~{\bf q}~,~\omega~-~E)$ are the 
spectral functions describing the energy and momentum distributions of the struck nucleon in the initial (hole) and 
final (particle) states, respectively.

The solid line of Fig. \ref{xsec_ee} represents the results
of a theoretical calculation of the QE contribution, carried out within the IA
using the hole spectral function of Ref.\cite{PkE} and the recent parametrization of the vector form factors 
of Ref. \cite{bodek}. Final state interactions (FSI) between
the struck nucleon and the recoiling  spectator system \cite{PRD}, whose main effect is a~$\sim 10$ MeV shift of 
the QE peak, have been also taken into account. 

It is apparent that height, position and width of the QE peak, mostly driven by the energy and momentum dependence of the 
hole spectral function, are well reproduced. 
 
\begin{figure}[ht]
\includegraphics[width=0.80\columnwidth]{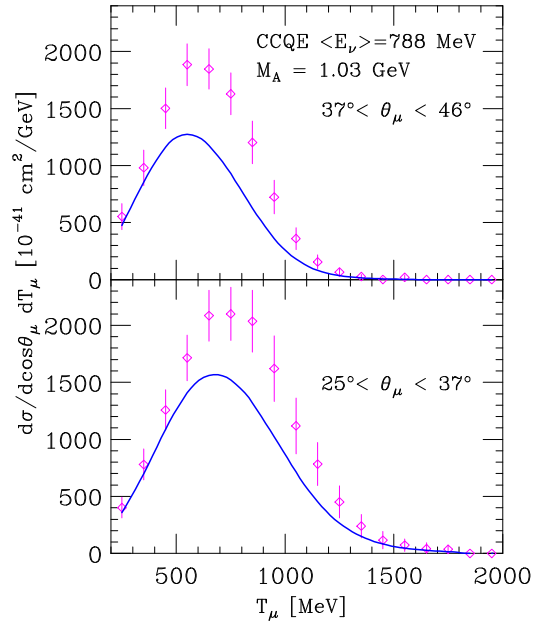}
\caption{(color online). Flux integrated double differential CCQE cross section measured by the MiniBooNE collaboration 
\cite{BooNECCQE},  shown as a function of kinetic energy of the outgoing muon. The upper and lower panels correspond to  
to different values of the muon scattering angle. Theoretical results have been obtained using the same spectral 
functions and  vector form factors employed in the calculation of the electron scattering cross section of Fig. \ref{xsec_ee}, 
and a dipole parametrizaition of the axial form factor with $M_A=1.03$ MeV.}
\label{dsigma}
\end{figure}

Applying the same scheme employed to obtain the solid line of Fig. \ref{xsec_ee} to neutrino scattering 
one gets the results shown in Fig. \ref{dsigma}. The data points represent the double differential CCQE cross 
section integrated over the MiniBooNE neutrino flux, whose average energy is \
$\langle~E_\nu~\rangle~=~788$~MeV, 
plotted as a function of the kinetic energy of the outgoing muon at different 
values of the muon scattering angle. 
The solid lines show the results (integrated over the $\cos \theta_\mu$ bins) obtained using the same spectral 
functions and  vector form factors employed in the calculation of the electron scattering cross section of Fig. \ref{xsec_ee}, 
and a dipole parametrization of the axial form factor with $M_A=1.03$~MeV. Comparison of Figs. \ref{xsec_ee} and 
\ref{dsigma} indicates that the electron and neutrino cross sections corresponding to the same target and 
comparable kinematical conditions (the position of the QE peak in Fig. \ref{xsec_ee} corresponds to 
kinetic energy of the scattered electron $\sim 610$ MeV) cannot be explained using the same theoretical approach
using the value of the axial mass resulting from deuterium measurements. In this instance, the paradigm of electron 
scattering appears to conspicuously fail. 

Note that the comparative analysis of electron and neutrino data, based 
on double differential cross sections depending on {\em measured} kinematical variables, is made possible for the 
first time by the availability of the data set of Ref. \cite{BooNECCQE}.  

The results of a global comparison between the MiniBooNE data and the calculated cross sections 
show that theory sizably underestimates the measured cross section for {\em any} values of muon energy and 
scattering angle.

It has to be emphasized that the above conclusion, while being based on a calculation carried 
out within the scheme of Refs. \cite{PRD,NPA}, is largely model independent. Theoretical approaches providing 
a quantitative description of the electron-nucleus cross section in the QE channel, are bound to predict CCQE
neutrino-nucleus cross sections significantly below the MiniBooNE data if the value of the axial mass is set to
1.03 GeV. 

\begin{figure}[ht]
\includegraphics[width=0.80\columnwidth]{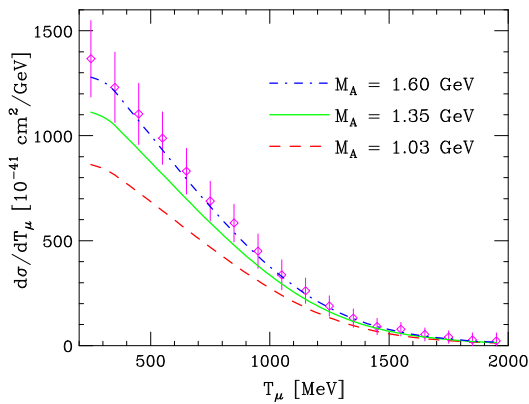}
\caption{(color online). Flux integrated muon kinetic energy spectrum. The dot-dash, solid and dashed lines 
have been obtained setting the value of the axial mass to $M_A = 1.03, \ 1.35 \ {\rm and} \  1.6$ GeV, respectively.
The data are taken from Ref.\cite{BooNECCQE} . }
\label{Tdist}
\end{figure}

In spite of the fact that the large value of $M_A$ reported by the MiniBooNE collaboration was obtained from 
a shape analysis of the $Q^2$-distribution, the effect of the axial mass on the CCQE cross section can be best 
analyzed studying the flux integrated muon kinetic energy spectrum and angular distribution, obtained integrating 
the double differential cross section over $\cos \theta_\mu$ and $T_\mu$, respectively. These quantities only 
depend on the {\em measured} muon kinematical variables, thus being unaffected by the assumptions 
associated with the reconstruction of the incoming neutrino energy $E_\nu$  \cite{shape,axmass}, entering the definition 
of the reconstructed $Q^2$.

Figure \ref{Tdist} shows a comparison between the MiniBooNE flux integrated muon kinetic energy spectrum and the results of our calculations, corresponding to three different values of $M_A$.  The behavior of the curve corresponding to $M_A = 1.03$ GeV is 
consistent  
with that shown in Fig. \ref{dsigma}, as the data turns out to be largely underestimated. 
Increasing $M_A$ to 1.35 GeV, the value resulting from the MiniBooNE analysis of Ref.\cite{BooNECCQE}, while improving the agreement between 
theory and experiment, still does not lead to reproduce the data at $T_\mu \lsim$ 1 GeV. The dot-dash curve has been 
obtained using the value $M_A = 1.6$~GeV, yielding the best $\chi^2$-fit to the MiniBooNE flux integrated 
$Q^2$-distribution within the our approach.

\begin{figure}[bht]
\includegraphics[width=0.80\columnwidth]{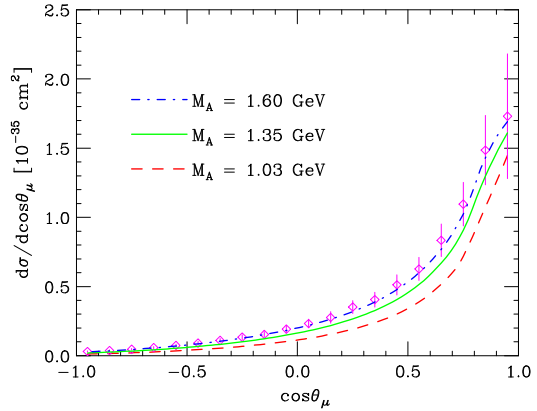}
\caption{(color online).  Flux integrated muon angular distribution. The dot-dash, solid and dashed lines 
have been obtained setting the value of the axial mass to $M_A = 1.03, \ 1.35 \ {\rm and} \ 1.6$ GeV, respectively.
The data are taken from Ref.\cite{BooNECCQE} .}
\label{angdist}
\end{figure}

The $M_A$-dependence of the flux integrated muon angular distribution, is shown in Fig. \ref{angdist}, together 
withe data from Ref. \cite{BooNECCQE}. The overall picture is clearly the same as in Fig. \ref{Tdist}.

\begin{figure}[bht]
\includegraphics[width=0.75\columnwidth]{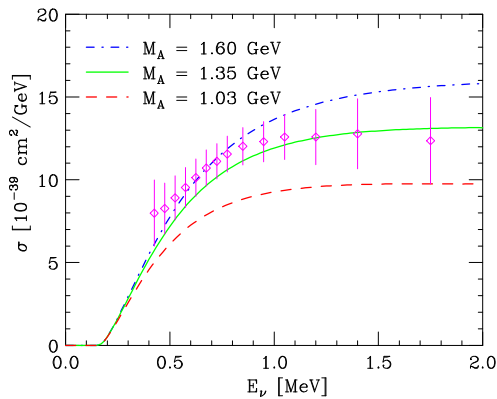}
\caption{(color online).  Flux unfolded total CCQE cross section, as a function of neutrino energy. 
The dot-dash, solid and dashed lines have been obtained setting the value of the axial mass 
to $M_A = 1.03, \ 1.35 \ {\rm and}  \ 1.6$ GeV, respectively. The data are taken from Ref.\cite{BooNECCQE} .}
\label{sigmatot}
\end{figure}

In Fig. \ref{sigmatot} we compare the results of our calculations to the MiniBooNE flux unfolded total cross section. 
It is apparent that in this case using $M_A=1.6$ GeV leads to overestimating the data in the region of high energy 
($E_\nu > 800$ MeV), 
where the choice  $M_A=1.35$ GeV provides a better fit. The different pattern emerging from Fig. \ref{sigmatot}, compared 
to Figs. \ref{Tdist} and \ref{angdist}, clearly points to the uncertainty associated with the interpretation of flux averaged and 
flux unfolded data. 

The results of our work indicate that the theoretical approach based on the IA and realistic spectral functions, 
successfully applied to QE electron scattering, fails to reproduce the CCQE  neutrino-nucleus cross section, 
unless the value of the nucleon axial mass resulting from deuteron measurements is modified.
In addition, the possibility of interpreting the large $M_A$ resulting from the K2K and MiniBooNE analyses
as an effective axial mass, modified by nuclear effects beyond the Fermi gas model, appears to be ruled 
out \cite{axmass}.
This statement should be regarded as largely model independent, as calculations carried out using  
different descriptions of nuclear dynamics yield similar results \cite{butke}. 

A different scenario is suggested by the results of Ref.~\cite{martini}, whose authors obtain a 
quantitative account of the MiniBooNE flux unfolded total cross section {\em without increasing} $M_A$. 
Within the model of Ref.~\cite{martini}, the main mechanism responsible for the enhancement that brings 
the theoretical cross section into agreement with the data is multi nucleon knock out,  leading to 
two particle-two hole (2p-2h) nuclear final states. 

Within the approach employed in our work, the occurrence of 2p-2h final states is described by the continuum part of
 the spectral function, arising from nucleon-nucleon correlations \cite{PkE}. It gives rise to the tail 
extending at large $\omega$, clearly visible in Fig. \ref{xsec_ee}. However, its contribution 
turns out to be quite small (less than 10\% of the integrated spectrum of Fig. \ref{xsec_ee}).  
According to the philosophy outlined in this paper, in order to firmly establish the role of multi-nucleon 
knock out in CCQE neutrino interactions the model of  
Ref.~\cite{martini} should be thoroughly tested against electron scattering data.


In our opinion, 
the available theoretical and experimental information suggests that the main 
difference involved in the analysis of neutrino-nucleus scattering, as compared to electron-nucleus scattering, lies in the 
flux average. 

Unlike the electron cross section shown in Fig. \ref{xsec_ee}, the flux averaged CCQE neutrino cross section 
at fixed energy and scattering angle of the outgoing lepton picks up contributions from different kinematical regions, where 
different reaction mechanisms 
dominate. As a consequence, it cannot be described according to the paradigm successfully 
applied to electron scattering, based on the tenet that the lepton kinematics is fully determined.

A {\em new paradigm}, suitable for the studies of neutrino interactions, should be based on a more flexible model of 
nuclear effects, providing a consistent description of the broad kinematical range corresponding to the relevant neutrino 
energies. 

Besides single- and multi-nucleon knock out, such a model should include the contributions of processes involving the 
nuclear two-body currents, which are known to provide a  significant enhancement of the electromagnetic nuclear 
response in the transverse channel \cite{euclidean}.  In addition, the occurrence of processes leading to pion 
production and excitation of nucleon resonances should also be taken into account. 

A great deal of information could be obtained applying the new paradigm to the analysis of {\em inclusive} data, 
preferably, although not necessarily, through direct implementation in the Monte Carlo simulation codes. This kind
of analysis may help to reconcile the different values of the axial mass obtained from different experiments, and would 
be unaffected by the problem of the possible misidentification of CCQE events, recently 
discussed in Ref. \cite{leitner}.

The authors are indebted to R. Schiavilla for a critical reading of this manuscript. Useful discussions 
with G.T. Garvey are also gratefully acknowledged.


\begin{thebibliography}{99}

\bibitem{BooNECCQE}
A.A. Aguilar-Arevalo {\it et al.} (MiniBooNE Collaboration), Phys. Rev. D {\bf 81}, 092005 (2010). 

\bibitem{VFF}
C.F. Perdrisat, V. Punjabi and M. Vanderh\"aghen, Prog. Part. Nucl. Phys.
{\bf  59}, 694 (2007). 
 
\bibitem{bernard}
V. Bernard {\em et al.}, J. Phys. G {\bf 28}, R1 (2002).

\bibitem{bodek}
A. Bodek, S. Avvakumov, R. Bradford, and H. Budd, Eur. Phys .J. C {\bf 53}, (2008).

\bibitem{nomad}
V. Lyubushkin et al. (NOMAD Collaboration), Eur. Phys. J. C {\bf 63}, 355 (2009).

\bibitem{K2K}
R. Gran {\em et al.} (K2K Collaboration), Phys. Rev. D {\bf 74},
 052002 (2006).

\bibitem{BOONE} A.A. Aguilar Arevalo {\em et al.} (MiniBooNE Collaboration), Phys. Rev.
Lett. {\bf 100}, 032301 (2008).

\bibitem{nuint09}
Proceedings of the {\em Sixth International Workshop on Neutrino-Nucleus Interactions
in the Few-GeV Region (NUINT-09)}. Eds. F. Sanchez, M. Sorel and L. Alvarez-Ruso.
AIP Conference Proceedings,  Vol. 1189 (2010).
 
\bibitem{martini}
M. Martini, M. Ericson, G. Chanfray and  J. Marteau, Phys. Rev.C {\bf 80}, 065501 (2009);
{\em ibidem} {\bf 81}, 045502 (2010).

\bibitem{RMP}
O. Benhar, D. Day and I. Sick, Rev. Mod. Phys. {\bf 80}, 189 (2008).

\bibitem{PRD}
O. Benhar, N. Farina, H. Nakamura, M. Sakuda and R. Seki, Phys. Rev. D {\bf 72}, 053005 (2005).

\bibitem{NPA}
O. Benhar and D. Meloni, Nucl. Phys. {\bf A789}, 379 (2007).

\bibitem{12C_ee}
J. S. O$^\prime$Connell {\em et al.}, Phys. Rev. C {\bf 35}, 1063 (1987).

\bibitem{PkE}
O. Benhar, A. Fabrocini, S. Fantoni and I. Sick, Nucl. Phys. {\bf A579}, 493 (1994).

\bibitem{NUINT07}
O. Benhar, AIP Conference Proceedings,  Vol. 967,  111 (2007).

\bibitem{shape}
A. Ankowski, O. Benhar and N. Farina, arXiv:1001.0481 [nucl-th].

\bibitem{axmass}
O. Benhar and D. Meloni, Phys. Rev. D {\bf 80}, 073003 (2009).

\bibitem{butke}
A. V. Butkevich, arXiv:1006.1595v1 [nucl-th].

\bibitem{euclidean}
J. Carlson and R. Schiavilla, Rev. Mod. Phys. {\bf 70}, 743 (1998)

\bibitem{leitner}
T. Leitner and U. Mosel, arXiv:1004.4433 [nucl-th].

\end{thebibliography}
\end{document}